\documentclass[twocolumn]{svjour3}         % twocolumn

\usepackage{graphicx,hyperref}
\usepackage{amsmath,bbm}
\newcommand{\rmd}{\mathrm{d}}
\DeclareMathOperator{\sech}{sech}
\DeclareMathOperator{\csch}{csch}
\DeclareMathOperator{\sinc}{sinc}
\newcommand{\av}[1]{\overline{#1}}

\begin{document}

\title{Localization of solitons: linear response of the mean-field ground state to weak external potentials}
\titlerunning{Localization of solitons}    

\author{Cord A.\ M\"uller}
\institute{C.~A.~M\"uller, Centre for Quantum Technologies,  National University of Singapore, 
  Singapore 117543, Singapore\\
  \email{cord.mueller@nus.edu.sg}}

\date{\today}

\maketitle

\begin{abstract}
Two aspects of bright  matter-wave solitons in weak external
potentials are discussed. First, we briefly review recent results on
the Anderson localization of an entire soliton in disordered potentials \cite{Sacha2009}, as a
paradigmatic showcase of genuine quantum dynamics beyond simple
perturbation theory. 
Second, we calculate the linear response of the mean-field soliton shape
to a weak, but otherwise arbitrary external potential, with a detailed
application to lattice potentials. 
\end{abstract}
\PACS{03.75.Lm 	Tunneling, Josephson effect, Bose-Einstein condensates
  in periodic potentials, solitons, vortices, and topological
  excitations 
\and 05.60.Gg 	Quantum transport}

%-------------------------------------------------------------------
\section{Introduction}
\label{intro.sec}

Recently, the cold-atom community has shown renewed interest in soliton dynamics, sparked by the 
experimental observation of 
cold-atom solitons in quasi-onedimen\-sional Bose-Einstein condensates
with attractive contact interaction
\cite{Khaykovich2002,Strecker2002}. 
Notably, it has been emphasized that the soliton's center of mass
is a collective degree of freedom whose dynamics can show
genuine quantum effects. In this
vein, Weiss and Castin 
\cite{Weiss2009} have calculated the scattering amplitude of a soliton
by a potential barrier, which results in a superposition of classically distinct
quantum states, namely the soliton being either transmitted or
reflected. Similarly,  Lewenstein and Malomed 
\cite{Lewenstein2009} have proposed to generate
entanglement by the controlled collision of quantum solitons. 

A well-known paradigm of genuine quantum dynamics is disorder-induced
Anderson localization \cite{Lifshitz1988,Mueller2010}, which has been
studied for solitons in different settings some time ago \cite{Gredeskul1992}, and 
also been observed rather recently with ultracold, 
noninteracting matter waves \cite{Billy2008,Roati2008,Sanchez-Palencia2010,Modugno2010} (see also
\cite{Albert2010}). 
Motivated by these experimental advances, 
we have investigated the quantum dynamics of matter-wave solitons in
spatially correlated disorder potentials \cite{Sacha2009}. 
The first part of the present paper reviews briefly
the derivation of an effective Hamiltonian for the center of mass and
the resulting localization exponent in an optical speckle potential.  

Naturally, there is more to solitons than just their center-of-mass
dynamics. Whenever inhomogeneous force fields act on a compound
object, the latter responds by adapting its internal configuration as well. For weak forces, the
response will be linear and thus described by a (linear)
susceptibility. The second part of this paper investigates in more
detail how the soliton's ground-state shape changes under the influence of a weak
external potential. We calculate the linear compressibility in
general, and then focus on the simple, yet interesting case of a lattice
potential, including a detailed comparison of analytical results to
numerical data. This allows us finally
to derive quantitative criteria for the external perturbation of the
shape to be weak.

%-------------------------------------------------------------------
\section{Setting the stage}
\label{stage.sec}

We describe a weakly interacting Bose-Einstein condensate (BEC) in a
quasi-onedimensional wave guide  
by its mean-field amplitude $\phi(z)$.  The Gross-Pitaevskii (GP) free
energy functional, with a given chemical potential $\mu$ and in a
homogeneous wave guide, reads  
\begin{equation} \label{E0.eq}
E_0[\phi,\phi^*] = \int \rmd z \left\{
\frac{1}{2}|\partial_z\phi|^2 + \frac{g}{2}|\phi|^4 - \mu |\phi|^2 
\right\}. 
\end{equation}
We will use units such that $\hbar=m=1$ throughout the following. 
$g=2\omega_\perp a$ is the effective interaction constant for a
quasi-onedimensional condensate with s-wave scattering length $a$ and
transverse harmonic trapping frequency $\omega_\perp$.  
The cases of repulsive ($a>0$) and attractive ($a<0$) 
interaction correspond to  $g>0$ and $g<0$, respectively. 

Minimizing the free energy $E_0$ yields the ground state
$\phi_0(z)$.  The chemical potential
$\mu$ thus determines the total number of particles $N_0=\int \rmd z
n_0(z)$, with $n_0(z)=|\phi_0(z)|^2$ being the condensate density.
Let us choose periodic boundary conditions, as for a toroidal wave
guide with circumference $L$. In the case of repulsive interaction
$g>0$, both kinetic energy and interaction energy are mimimized by
spreading the density homogeneously over the entire available length:
$n_0 = \mu/g$ and $N_0=L\mu/g$ with $\mu>0$.  
The corresponding ground-state wave function $\phi_0= e^{i\theta_0}
\sqrt{\mu/g}$ plays the r\^ole of a BEC order parameter,  
featuring an arbitrary global phase $\theta_0$ that spontaneously breaks the
$U(1)$ gauge invariance of \eqref{E0.eq}. 
For the ground state of a
single Bose-Einstein condensate, we can
set $\theta_0=0$.

An attractive interaction $g<0$ rather favors a state where atoms are
clustered together. As shown by Kana\-moto et al.\ \cite{Kanamoto2003}, 
for large enough system size or chemical potential
(i.e. number of particles), $L|\mu|^{1/2}\gg1$, the energetically
preferred state is a \emph{soliton},   
\begin{equation} \label{phi0.eq}
\phi_0(z-z_0) = \left|\frac{2\mu}{g}\right|^{1/2}\sech\left[(z-z_0)/\xi\right] e^{i\theta_0}
\end{equation}
with a hyperbolic secant envelope, decaying over the length
scale $\xi=|2\mu|^{-1/2}$ known as the condensate healing length. 
The number of particles is now $N_0=2|2\mu|^{1/2}/|g|$. Conversely,
in a canonical setting with fixed number of particles $N$, the chemical
potential settles to $\mu_0=-1/(2\xi^2) = - g^2N^2/8 <0$. 

Note that the solution \eqref{phi0.eq} spontaneously breaks not only the gauge invariance
with a phase $\theta_0$ (that we take to be zero), 
but also the translational invariance of the energy functional
\eqref{E0.eq}.  Therefore, the center of mass $z_0=:q$ emerges as a
dynamical degree of freedom on its own. 
In the homogeneous situation described by \eqref{E0.eq}, all classical
solutions $q(t)=q_0+v t$ with constant velocity $v$ are admissible 
by Galilean invariance.
In a given frame of reference, the classical configuration with
minimum kinetic energy is a soliton  resting at $q=q_0$. Quantum mechanically,
however, $q$ is rather distributed with its ground-state wave
function $\Psi_0(q)$ that   
obeys the free Schr\"odinger equation under the given boundary conditions. 
For periodic boundary conditions, this center-of-mass ground state is the completely 
delocalized plane wave with momentum $p_q=0$, i.e. the constant $\Psi_0(q) =
L^{-1/2}$. 
Contrary to the point of view that the ``wave properties of solitons manifest themselves in radiation
emitted due to scattering by impurities'' \cite[p.53]{Gredeskul1992}, we take a more
general approach along the fundamental principles of quantum
mechanics, namely that wave-particle duality is a general feature of
all degrees of freedom, to be revealed under appropriate
experimental circumstances. 

It may seem peculiar to speak of quantum dynamics for the center of mass of a whole collection of atoms, 
after starting out from a mean-field description for the
entire condensate. 
But it is indeed very common to separate the center of mass from
internal variables in 
interacting systems and moreover exact for two-body forces that only
depend on the relative distance between the microscopic constituents.
The situation of a soliton, composed of individual atoms 
held together by attractive contact interactions, is therefore
analogous to the situation of, say, an alkali atom composed 
of a nucleus and electrons held together by the Coulomb force. 
And just as one may study the quantum dynamics of entire atoms
---not to mention fullerenes or biomolecules
\cite{Hackermuller2003}---one
may indeed equally well study the quantum dynamics of entire
solitons as far as their center of mass is concerned. 

Consider now the mean-field energy functional in presence of an external potential 
$V(z)$: 
\begin{equation} \label{E.eq}
E= \int \rmd z \left\{
\frac{1}{2}|\partial_z\phi|^2 + \frac{g}{2}|\phi|^4 + [V(z) - \mu] |\phi|^2 
\right\}.   
\end{equation}
We assume that $V(z)$ is a small perturbation on the energy scale set
by $\mu$.
To lowest order in $V/|\mu|$, therefore, the shape of the original
soliton will remain unchanged. 
In contrast, the external potential can never 
be considered a small perturbation for the center-of-mass dynamics  
because a finite $V$ is never small compared to the homogeneous case
$V=0$. 

At this point, the paper bifurcates. In the following
section, we focus on the center of mass. We highlight the quantumness
of its dynamics in a disordered potential by discussing the Anderson localization
length, as first derived in \cite{Sacha2009}. 
As a complement to this work, we will study  in sections  \ref{linearResponse.sec} and \ref{lattice.sec}
below how the soliton's shape is modified by
the presence of a weak external potential in the mean-field ground state. Section
\ref{conclusion.sec} concludes.

%-------------------------------------------------------------------
\section{Anderson localization of a soliton}
\label{locSoliton.sec}

%-------------------------------------
\subsection{Effective Hamiltonian} 

Assuming a fixed soliton shape, the center of mass $q$ of $N$
particles can be described as a collective
variable with the ansatz 
$\phi(z;q,p_q) =  e^{i p_q z /N}\phi_0(z-q)$.  Here, the conjugate momentum $p_q$ 
appears in the phase together with a factor $N^{-1}$ because $\phi_0$,
eq.~\eqref{phi0.eq}, is normalized to $N=2|2\mu|^{1/2}/|g|$.    
Inserting this collective-variable ansatz in the energy functional
\eqref{E.eq} and integrating over $z$  
yields the effective Hamiltonian
\begin{equation} \label{Hq.eg}
H_q = \frac{p_q^2}{2N} + \int \rmd z |\phi_0(z-q)|^2 V(z).  
\end{equation} 
This Hamiltonian describes a particle with mass $N$ evolving in a
potential $\widetilde V(q)= V\ast n_0(q)$ that is the convolution of the bare
potential with the soliton envelope.  

If the bare potential varies
only very slowly over one healing length $\xi= 2/(N|g|)$, then the soliton feels
the sum of forces on its constituents, $\widetilde V(q) = NV(q)$. If on
the other hand the potential varies rather rapidly, then the
convolution by the soliton density washes out all details on scales
smaller than $\xi$, and the effective potential is strongly reduced. This is
easily illustrated with a lattice potential $V(z) = V_0\cos(k z)$, for
which the effective potential is essentially the Fourier transform of
the soliton density $|\phi_0(z-z_0)|^2$ of \eqref{phi0.eq}:  
\begin{equation}\label{tildeVq.eq}
\widetilde V(q) = N \frac{\pi k\xi/2}{\sinh(\pi k\xi /2)} V_0 \cos(k
q). %= \widetilde{V}_0 \cos(kq). 
\end{equation} 
Indeed, $\widetilde V(q) \to  N V(q)$ as $k\xi\to 0 $, and for $k\xi
\gg 1$,  the amplitude  
$\widetilde{V}_0 \sim k e^{-\pi k \xi /2}$ becomes exponentially
small.

%-----------------------------------
\subsection{Correlated disorder}

A disorder potential is a random process $V(z)$ characterized by its
moments $\av{V(z)}$, $\av{V(z_1)V(z_2)}$, etc.  
Statistically homogeneous disorder is translation invariant after
averaging, with $\av{V(z)}=const.$, $\av{V(z+z_0)V(z_0)}= \av{V(z)V(0)}$,
etc. Without loss of generality, one can always set $\av{V(z)}=0$ by
redefining the zero of energy or shifting $\mu \mapsto \mu+\av{V(z)}$  in
\eqref{E.eq}. Thus, the most basic information about the disorder
potential is its pair correlator 
$\av{V(z)V(0)} = V_0^2 C(z/\sigma)$ where $V_0^2:= \av{V(z)^2}$ is the variance
characterizing the overall strength of disorder. The spatial
correlation function $C(z/\sigma)$ decreases from $C(0)=1$ to zero 
over a characteristic length scale $\sigma$,  the correlation length
of the disorder. 

Equivalently, a disordered potential can be
seen as a random superposition of plane-wave components
$V_k = L^{-1}\int\rmd z \, e^{-i k z} V(z)$. 
Statistical homogeneity then translates into conservation of
total momentum under averaging: 
$\av{V_{k}V_{k'}} =L^{-1}\delta_{k,-k'} V_0^2 P(k)$, 
where the so-called power spectrum $P(k)$ is the Fourier
transform of the real-space correlator $C(z/\sigma)$. 

For the present case of cold-atom dynamics, we
consider in detail optical speckle potentials \cite{Clement2006} 
for which the laws of
optics result in a remarkably simple correlation: 
$C(z/\sigma) = [\sinc(z/\sigma)]^2$ or 
\begin{equation}  \label{P_k.eq}
P(k) = \pi \sigma(1-\tfrac{1}{2}|k\sigma|)\Theta(1-\tfrac{1}{2}|k\sigma|). 
\end{equation} 
The correlation length $\sigma$ is determined by the wave length of the laser
light and the geometric aperture of the imaging system and can be as
short as $\sigma=0.26\,\mu$m 
\cite{Billy2008}. The Heaviside distribution
$\Theta(.)$ in \eqref{P_k.eq} excludes all wave vectors with modulus larger than
$2/\sigma$, as required by the limit of optical
resolution.  

The effective potential felt by the soliton's center of mass is the
convolution of the bare potential by the soliton density. By virtue of
the convolution theorem, the Fourier components of the effective
potential are therefore the
product of the bare components times the Fourier components of the
density, which already appeared in \eqref{tildeVq.eq}: 
\begin{equation}
\widetilde V_k= N \frac{\pi k\xi/2}{\sinh(\pi k\xi /2)} V_k. 
\end{equation}

The statistical properties of the potential affecting the soliton's
center of mass are therefore completely determined and readily
expressed in Fourier components. 
For example, the effective power spectrum reads 
\begin{equation}
\widetilde P(k) = N^2 \frac{(\pi k\xi/2)^2}{\sinh(\pi k\xi /2)^2} P(k). 
\end{equation}
If the  potential is very smooth on the soliton scale $\xi$, i.e. has
a correlation length $\sigma\gg\xi$, the power spectrum $P(k)$ goes to
zero faster than the soliton Fourier envelope, and $\widetilde P(k) \approx N^2 P(k)$. Conversely, if
the bare potential is varying very rapidly, $\sigma\ll\xi$, the bare
spectrum can be approximated in the range $k\sigma\ll1$ by its delta-correlation limit $P(0)$. 
Then, the healing length $\xi$ takes over as the new correlation length
with an exponential decay of potential fluctuations as  
\begin{equation}
\widetilde P(k) \approx (N \pi k\xi)^2 P(0) e^{-\pi k\xi}  
\end{equation}
for $k\xi\gg 1$.

%------------------------------------------------
\subsection{Anderson localization exponent}

The Hamiltonian of the free soliton, $p_q^2/2N$, has the
 plane-wave eigenfunctions $\Psi_k(q)\propto e^{i k q }$. When a small disorder
 potential is switched on, these extended plane-wave states become
 exponentially localized. Mathematically rigorous theorems assure that
 in 1D, the logarithmically averaged eigenfunctions decay for large
 distances $q$ from their origin like  \cite{Lifshitz1988,Gredeskul1992}
\begin{equation} 
\lim_{q\to \infty}\av{\log |\Psi(q)|} = - \frac{1}{2} \gamma(k)
 q. 
\end{equation} 
 The corresponding localization exponent, or inverse localization
 length, $\gamma(k)$ can be calculated perturbatively (for generic
 values of $k$, i.e., away from the band center $k=0$ or other
 singular points) in powers of the strength of the disorder
 potential. To second order in $V_0$, in the so-called Born approximation, 
the localization exponent reads
$
\gamma(k) = (k^2V_0^2/4E_k^2) P(2k)$, 
with $E_k=\hbar^2 k^2/2m$ the free kinetic energy 
\cite{Sanchez-Palencia2007,Lugan2009,Gurevich2009}. 
 The potential correlator is evaluated at momentum $2k$
since it is the elementary backscattering process $k\to -k$ that 
is eventually responsible for Anderson localization in 1D. 

%---------------------------------------------
\begin{figure}
\begin{center}
\includegraphics[width=0.9\linewidth]{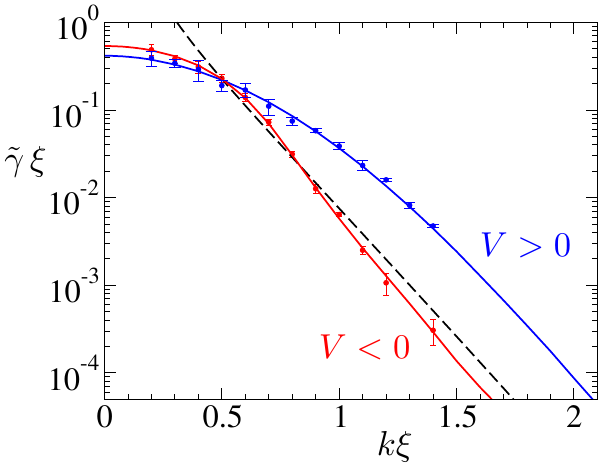}
\caption{Log-linear plot of the soliton localization exponent
  $\tilde\gamma(k)$ versus
  its center-of-mass wave vector $k$, in units of the soliton width
  $\xi$. Circles are numerical results obtained by diagonalizing the
  Hamiltonian  \eqref{Hq.eg} and solid lines are the result of a
  transfer-matrix calculation \cite{Sacha2009,Sacha2009a}, for  a red- and blue-detuned speckle
  potential with $V_0= \pm
  8\cdot10^{-5}|\mu|$, respectively, and a correlation length $\sigma=0.28\xi\approx
  0.26\,\mu$m. The Born approximation, Eq.\
  \eqref{gammakSoliton.eq}, is shown as a dashed line. The
  overall exponential decrease for $k\xi\gg 1$ and $k\sigma\ll1$ is
  clearly visible.  
}
\label{gammak.fig}
\end{center}
\end{figure}
%-----------------------------------------------------

For the soliton, we have $\widetilde{E}_k=k^2/(2N)$ and scattering by the effective
potential $\widetilde V_k$, such that the localization exponent is predicted to be 
\begin{equation} \label{gammakSoliton.eq}
\tilde\gamma(k) = \frac{N^2 V_0^2}{ k^2} 
\widetilde P(2k) 
= \frac{N^4 V_0^2}{ k^2} \frac{(\pi k\xi)^2}{\sinh(\pi k\xi
  )^2} P(2k). 
\end{equation} 
Figure \ref{gammak.fig} shows this prediction, together with 
numerical data, obtained both by exact diagonalization of the
Hamiltonian \eqref{Hq.eg} and a transfer matrix approach,
respectively \cite{Sacha2009,Sacha2009a}. We have chosen realistic experimental parameters:
$N=100$ ${}^7$Li atoms with scattering length $a=-3\,$nm in a
transverse trap with $\omega_\perp= 2\pi\times 5\,$kHz form a soliton
of size $\xi\approx 100\,\mu$m$/N\approx1\,\mu$m. We consider an optical speckle
potential with amplitude $V_0= \pm 8\cdot10^{-5}|\mu|$, i.e. both the
red-detuned case with $V_0<0$ and the blue-detuned case with
$V_0>0$. Since the speckle potential has a non-Gaussian, skewed
distribution, the full localization exponent depends on the absolute
sign of $V_0$, an effect that the lowest-order Born approximation
$O(V_0^2)$ cannot capture. However, the overall exponential decrease for
$k\xi\gg1$ is correctly predicted.

These results show that the localization length, $\tilde\gamma^{-1}$, 
of not-too-fast solitons with $k\xi\approx 1$ is in the sub-mm range
that is measurable in current experiments on 
localization of non-interacting matter waves \cite{Billy2008}. Further
results  can be found in \cite{Sacha2009}, such as the expected
density distribution of the final soliton position, depending on
initial trapping conditions.

But what about the emission of radiation due to scattering by
impurities, as discussed at length by Gredeskul and Kivshar
\cite{Gredeskul1992} and others? It is important to realize that
we specifically focus on smooth,
spatially correlated potentials that are expected to provoke
considerably less excitation than isolated, $\delta$-like
impurities. Nonetheless, it is clearly important to know the precise 
effects of the external disorder on the soliton shape. Some light on
this issue will be shed in the following sections.

%-------------------------------------------------------------------
\section{Linear response for the mean-field ground state}
\label{linearResponse.sec}

The previous results on the disorder-induced localization of a soliton
have been derived with the effective Hamiltonian \eqref{Hq.eg} as the
only ingredient. There, the soliton shape was assumed to be completely
unaffected by the external potential. Although $|V| \ll |\mu| $ is
probably a sufficient condition for this to be valid 
\cite{Sacha2009a}, as such it
cannot be a satisfying, namely necessary criterion. Indeed,
a potential with a large amplitude $V\gg|\mu|$ but fluctuating 
only on very long length scales hardly
deformes the soliton if its variation over the soliton size, $\xi
\rmd V/\rmd z$,  is very small. Conversely, the amplitude of an 
interacting condensate cannot follow 
very rapid potential
oscillations with a wavelength much smaller than $\xi$ since this would cost too much kinetic
energy in \eqref{F.eq}. In this case, only strongly smoothed
fluctuations are expected to appear, just as in the repulsive case 
\cite{Sanchez-Palencia2006}. In order to sharpen the picture, 
we therefore set out to compute the mean-field soliton deformation as
function of the potential's amplitude and wave vector. 

A small deformation of the ground state can be described by an expansion over a
complete set of elementary excitations, such as described by Bogoliubov
theory. In general, the Bogoliubov theory of
condensate excitations in inhomogeneous potentials can be formulated
as a saddle-point expansion of the mean-field energy functional 
\cite{Gaul2008,Gaul2010} and comprises two
important steps: In a first step, the
deformed ground state is determined as a functional of the external
potential. In a second step, the quadratic excitations around this
deformed ground
state are determined. In order to arrive at meaningful results  with 
all effects to a given order of $V$ taken properly into account, the
first step is a vital prerequisite for carrying out the second. 
In the present contribution, we shall content
ourselves with taking step 1, the consistent calculation of the 
mean-field ground state deformation caused by a weak, but otherwise
arbitrary, external potential $V(z)$.  

For the repulsive case with its homogeneous ground-state density
$n_0=\mu/g$, such a linear response to an external inhomogeneous
potential can be computed rather easily, taking advantage of Fourier
decomposition; for applications to disordered potentials see, e.g.,
the work of Giorgini et al.\ \cite{Giorgini1994} and Sanchez-Palencia \cite{Sanchez-Palencia2006}. 
For a soliton with its inhomogeneous ground
state density, a very similar
linear response scheme, technically slightly more demanding but
perhaps also
more interesting, is
presented in the following.

%-------------------------------------------------------
\subsection{Density response} 
\label{suscep.sec}

Taking our cue from Giorgini et al., we start with the GP functional \eqref{E0.eq}
in density-phase representation $\phi(z)=\sqrt{n(z)}e^{i \theta(z)}$: 
\begin{equation} \label{F.eq}
E_0  = \int \rmd z \left\{
\frac{1}{2}\left[ \left(\partial_z\sqrt{n}\right)^2 +
  n(\partial_z\theta)^2\right] + \frac{g}{2}n^2 - \mu n 
\right\}.   
\end{equation}
Since the perturbation $\int \rmd V(z) n(z)$ couples to
 the density, the deformed ground-state density reads $n(z)= n_0(z) +
 \delta n(z)$, where the shift within linear response is given by 
\begin{equation} \label{density_response.eq}
\delta n(z) = -\int \rmd z' \chi(z,z') V(z') + O(V^2). 
\end{equation} 
The density-density susceptibility (essentially the compressibility) is defined
via its functional inverse, 
\begin{align} 
\chi^{-1}(z,z') & = \left.\frac{\delta^2 E_0}{\delta n(z)\delta
    n(z')}\right|_0 \\ 
&= \frac{1}{2}\left[ \frac{1}{2n_0(z)} \partial_z\partial_{z'} -
  \frac{\mu}{n_0(z)} + 3 g \right] \delta(z-z'). 
\end{align} 
Unlike in the repulsive case with $n_0$ constant where the inverse is easily
computed in Fourier modes \cite{Giorgini1994}, here we have an
inhomogeneous density $n_0(z)$
%=
%(2\mu/g)\sech[(z-z_0)/\xi]^2$ 
and thus $\chi^{-1}$ difficult
to invert. We bypass this difficulty with a second approach in the
upcoming section and come back to the compressibility in section \ref{realkernel.sec}
below.  

%--------------------------------------------------------
\subsection{Wave-function response} 

The mean-field ground-state wave function $\phi(z)$, which can be
taken real, satisfies $\delta E/\delta\phi^*=0$, known as  
the stationary GP equation
\begin{equation} 
\left[-\frac{1}{2} \partial_z^2 + g\phi(z)^2 - \mu + V(z)\right]
\phi(z)=0. 
\end{equation} 
Following Sanchez-Palencia \cite{Sanchez-Palencia2006}, we develop
$\phi(z)= \phi_0(z-z_0)+\delta\phi(z-z_0)$ around the ground-state solution
\eqref{phi0.eq} by treating $\delta\phi$ as a small quantity of order
$V/|\mu|$. Linearizing gives
\begin{equation} 
\left[-\frac{1}{2} \partial_z^2 +3gn_0(z-z_0) - \mu \right]
\delta \phi(z-z_0)= - V(z)\phi_0(z-z_0). 
\end{equation} 
Measuring distances in units of $\xi$ around $z_0$, i.e. writing
$z-z_0=\xi x$ 
and dividing by $2|\mu|=\xi^{-2}$, one finds the 
linear equation
\begin{equation} \label{delphiPT.eq}
\left[-\frac{1}{2} \partial_x^2 - 3 \sech(x)^2 + \frac{1}{2} \right]
\delta \varphi(x)= - \frac{V(z_0
+\xi x)}{2|\mu|} \varphi_0(x)
\end{equation} 
for the dimensionless shift  
$\delta\varphi(x) =
\sqrt{\xi}\delta\phi(\xi x)$. 
This equation is of the form 
\begin{equation} 
 \left[H_0 + \frac{1}{2} \right]
\delta \varphi(x)= W(x)
\end{equation} 
with 
$H_0 = p^2/2 - 3 \sech(x)^2$ the Hamiltonian of the
$\sech^2$-potential well, the  well-known  P\"oschl-Teller potential 
\cite{Poschl1933}. On the right-hand side, one has an external source term  
\begin{equation}\label{W.eq} 
W(x)= - \varphi_0(x)\frac{V(z_0 +\xi x)}{2|\mu|}. 
\end{equation}  

Computing  $\delta \varphi$  then only requires to invert $H_0$. 
Fortunately for us, this is very simple to do because
all potentials of the form $-\frac{1}{2}\nu(\nu+1)\sech(x)^2$ with
integer $\nu$ are
supersymmetric partners of the free-particle case
$\nu=0$, and their eigenstates and eigenenergies are perfectly known; 
for a brief and pedagogical introduction to these issues, see
\cite{Boya1988}. The case $\nu=2$ of interest here is treated in
detail by Lekner \cite{Lekner2007}. $H_0$ admits two bound states, 
\begin{align}  
\psi_0(x) &= \frac{\sqrt{3}}{2}\sech(x)^2, \\
\psi_1(x) &= \sqrt{\frac{3}{2}} \sech x \tanh x 
\end{align}  
with eigenenergies $E_0 = -2$ and $E_1=-\frac{1}{2}$ (in units of $2|\mu|$), respectively, 
as well as scattering eigenstates
\begin{equation} 
\psi_k(x)= \frac{e^{i k x}}{[2\pi]^{1/2}}
\frac{k^2-2+3 \sech(x)^2 + 3 i k \tanh x}{[(1+k^2)(4+k^2)]^{1/2}}
\end{equation} 
with free kinetic eigenenergy $E_k= k^2/2$, $k\in \mathbbm{R}$. 
These eigenmodes  have appeared repeatedly in one form or another in
the soliton literature (cf.\ a recent paper by Castin
\cite{Castin2009} and works cited therein), but often without
reference to the underlying
supersymmetry of the P\"oschl-Teller potential. 

We can expand the deformation $\delta \varphi$ over this orthonormal basis
set of eigenfunctions, 
\begin{equation}
\delta\varphi(x) = \alpha_0 \psi_0(x) + \alpha_1 \psi_1(x) + \int\rmd k
\alpha_k \psi_k(x),  
\end{equation} 
and then determine the coefficients by projecting \eqref{delphiPT.eq}
onto each eigenfunction:   
\begin{equation} \label{alphaj.eq}
 \left(E_j+\tfrac{1}{2}\right) \alpha_j  = \int\rmd y \, \psi^*_j(y) W(y). 
\end{equation} 
Interestingly, the 
coefficient $\alpha_1$ of the first excited bound state $j=1$ remains undetermined because $E_1 = -1/2$
of $H_0$ is exactly compensated by the $+1/2$. In other words, this mode appears as a zero-energy
eigenmode of the linear response kernel \cite{Sacha2009,Sacha2009a,zeromode}. 
This has a 
simple, yet profound physical explanation:  because $\psi_1(x) =
-\sqrt{3/N}  \partial_x\varphi_0(x)$, the deformation 
$\varphi_0(x)+ \alpha_1\psi_1(x) = \varphi_0(x+\delta x) +O(V^2)$ simply shifts the center of mass by 
$\delta x= -\alpha_1 \sqrt{3/N}$. Thus we find, as argued in Sec.~\ref{intro.sec} above, that the soliton's
center of mass is an independent dynamical
variable that is influenced non-perturba\-tively by the
external potential.  

Let us then assume in the following that the soliton's center of mass has
reached a position such that the external potential does not
accelerate it anymore, namely that the right hand side of
\eqref{alphaj.eq} for $j=1$ vanishes. A sufficient condition for this
is that the external potential $V(z)$ is (locally) an even function
around the soliton position $z_0$. Then, the soliton shape
deformation  reads    
\begin{equation} \label{delphi.eq}
\delta\varphi(x) = \alpha_0 \psi_0(x) + \int\rmd k
\alpha_k \psi_k(x)   
\end{equation} 
with eq.\ \eqref{alphaj.eq} determining the coefficients $\alpha_0$
and $\alpha_k$ as linear functions of the external potential
$V(z)$, scaling by construction as $V/|\mu|$. In Sec.\
\ref{lattice.sec} below, we will study the scaling with the wave vector for
a simple lattice potential.

%--------------------------------------------------------
\subsection{Real-space response kernel} 
\label{realkernel.sec} 

By linearity, the soliton shape deformation \eqref{delphi.eq} 
can also be conveniently expressed as a functional of the potential
\eqref{W.eq}, 
\begin{equation} \label{delphik.eq} 
\delta\varphi(x) = \int \rmd y K(x,y) W(y),  
\end{equation} 
with a symmetric kernel given by 
\begin{equation} \label{kernel.def}
K(x,y) = -\frac{2}{3} \psi_0(x)\psi_0(y) +2 \int \rmd k
\frac{\psi_k(x)\psi_k^*(y)}{1+k^2}. 
\end{equation}  
The $k$-integral can be evaluated, leading after some 
algebra to the following closed-form expression for the real-space response
kernel: 
\begin{align} \label{kernel.eq}
K(x,y) & =  \frac{1}{4} \sech x \tanh x \sech y \tanh y  \\
& \quad \times \big[ \cosh 2x + \cosh 2y  - |\sinh 2x - \sinh 2
  y|  \nonumber  \\
& \qquad - 4 \csch x  \csch y  \sinh|x-y|   - 6 |x-y|  \big]. \nonumber 
\end{align}  
Equations \eqref{delphik.eq} and
\eqref{kernel.eq} together with \eqref{W.eq} provide a pr\^et-\`a-calculer expression of the
soliton shape as function of any given potential $V(z)$. 
While soliton deformation in external potentials is certainly not a
new topic and has been considered in a rather large number of different contexts (see e.g.\ \cite{Gredeskul1992} and
references therein),  to our knowledge this useful expression of the
linear-response kernel is new.  

This solution determines also the compressibility 
introduced in Sec.~\ref{suscep.sec}. From  
$n(z)=[\phi_0(z-z_0)+\delta\phi(z)]^2 = n_0(z-z_0)+\delta
n(z)+O(V^2)$ it follows that $\delta n(z) = 2 \phi_0(z-z_0) \delta
\phi(z)$, and we can read off from the previous expressions that the
density-density susceptibility of eq.\ \eqref{density_response.eq} is given by 
\begin{equation}\label{suscep.eq}
\chi(z,z') = 2\xi \phi_0(z-z_0) K\left(\frac{z-z_0}{\xi},\frac{z'-z_0}{\xi}\right) \phi_0(z'-z_0). 
\end{equation}
The compressibility $\chi(z,z')$ does not depend solely on $z-z'$, as
it would in homogeneous systems. Instead, it refers to a distinguished point, namely the soliton
position $z_0$, and is only invariant under simultaneous translation
of all 3 coordinates $z,z',z_0$.  

%------------------------------------------------------
\subsection{Grand-canonical vs.\  canonical deformation} 

The previous derivation used the grand-canonical setting at fixed
chemical potential $\mu$. The corresponding soliton
deformation $\delta\varphi(x)=:\delta\varphi(x)|_\mu$ and density
shift $\delta n(z) =:\delta n(z)|_\mu$ do not conserve the total
number of particles. The change in particle number to 
order $V/|\mu|$ reads 
$\delta N =   \int \rmd z \delta n(z) = -\int \int \rmd z \rmd z'
\chi(z,z') V(z')$. 
 
In order to calculate the soliton deformation at fixed number of
particles $N$, i.e. to
compensate $\delta N$, the chemical potential has to be
adjusted. Since $|\mu| = g^2N^2/8$, the required shift is  
 $\delta |\mu| = - 2|\mu| \delta N/N$. Inserting \eqref{suscep.eq}  
and performing the
integral over $z$ yields the relatively simple expression
\begin{equation}\label{deltamu.eq}
\frac{\delta|\mu|}{2|\mu|}  =  - \int \rmd y (1-y \tanh y) \sech(y)^2 \frac{V(z_0+\xi y)}{2|\mu|}.
\end{equation} 
Since the grand-canonical deformation $\delta\varphi|_\mu$ given by \eqref{delphik.eq}  is already of
order $V/|\mu|$, it is not affected by this shift to lowest order, but the original soliton
background is changed: $\varphi_0(x)|_N = \varphi_0(x)|_\mu +
\partial_{|\mu|}\varphi_0(x)\delta|\mu|$. 
Altogether, we find for the canonical soliton deformation at fixed
number of particles $N$ 
\begin{equation} \label{deltaphiN.eq}
\delta\varphi(x)|_N = \delta\varphi(x)|_\mu +
\frac{\delta|\mu|}{2|\mu|} (1- x \tanh x) \varphi_0(x)  
\end{equation} 
where $\delta\varphi(x)|_\mu$ is given by \eqref{delphik.eq},  
and $\delta |\mu|/2|\mu|$ by \eqref{deltamu.eq}, with  
$\mu=-g^2N^2/8$, $\xi=2/(N|g|)$ and 
$\varphi_0(x)= \sqrt{N/2}\times\sech x$ in all expressions. 
This formula, together with the kernel \eqref{kernel.eq}, constitutes
the main achievement of this work.

%-------------------------------------------------------------------
\section{Special case: Lattice potential}
\label{lattice.sec}

In order to illustrate the above results, we study in detail the case
of a simple lattice potential with reduced wavevector $k_0= k\xi$:   
\begin{equation} \label{Vlattice.eq} 
V(\xi x) = -V_0 \cos(k_0 x)
\end{equation}
with $V_0\ge 0$ such that the center of mass of a soliton prepared at
$z_0=0$ (modulo the lattice period) sits in a potential minimum, where it remains
classically. The limit $k_0\ll1$ reduces to the particular case of a purely harmonic
confinement studied recently by Castin \cite{Castin2009}. 
Since the soliton deformation $\delta\varphi$ is a
linear functional of $V$, more general potentials (such as disordered
ones) can be studied by applying the following results to their Fourier components. 

%-------------------------------------------------------------------
\subsection{Chemical potential shift}

First of all, the  chemical
potential shift \eqref{deltamu.eq}  evaluates to 
\begin{align} \label{deltamu_lattice.eq}
\frac{\delta|\mu|}{|\mu|}  
%& =  \frac{V_0}{|\mu|} \int \rmd y (1-y \tanh y) \sech^2(y)\cos k_0x \nonumber \\
 &  =  \frac{V_0}{|\mu|} \frac{(\pi k_0 /2)^2}{\sinh(\pi k_0/2)^2}
 \cosh\frac{\pi k_0}{2}. 
\end{align} 
As $k_0=k\xi \to 0$, one finds $\delta|\mu|=V_0$ as expected, since this
exactly compensates a global potential offset $-V_0$. Also not
surprisingly, 
 the shift vanishes as
$k_0\to\infty$ at fixed $V_0$  since the condensate
cannot follow these rapid variations. But it may perhaps come as a
surprise that the behaviour as function of $k_0$ described by
\eqref{deltamu_lattice.eq} is non-monotonic. 

%---------------------------------------------
\begin{figure}
\begin{center}
\includegraphics[width=\linewidth]{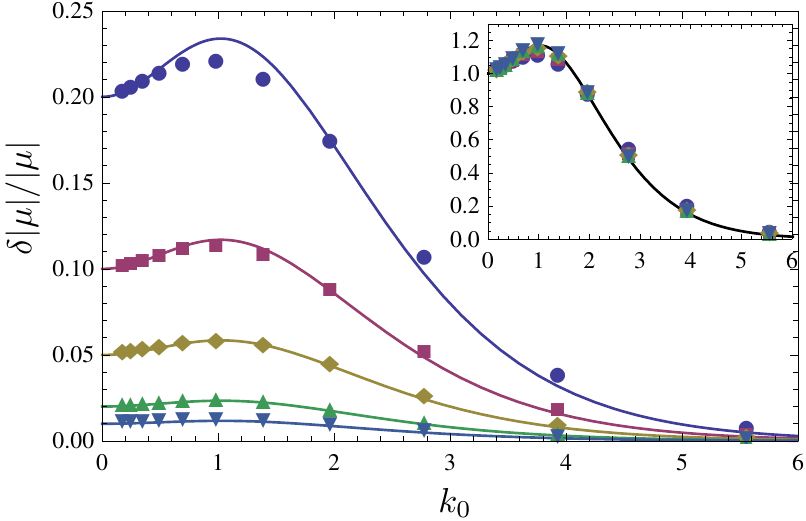}
\caption{(Color online) Relative shift in the chemical potential, $\delta|\mu|/|\mu|$
  as function of $k_0$ in a lattice of various depths
  $V_0/|\mu|\in\{0.2,0.1,0.05,0.02,0.01\}$ [top to bottom]. Symbols: Numerical results
from an imaginary-time integration of the GP equation. Solid lines:
linear-response prediction, eq.\ \eqref{deltamu_lattice.eq}. The inset shows
the collapsed data, i.e.\ the raw data divided by
$V_0/|\mu|$. As $k_0\gg1$ or $V_0/|\mu|\ll1$, the shift is correctly
predicted and, as expected, small.    
}
\label{mu_q.fig}
\end{center}
\end{figure}
%-----------------------------------------------------

We can compare these predictions to the results of a numerical integration of the 
imaginary-time GP equation
$\partial_\tau\varphi(x,\tau) = - H \varphi(x,\tau)$ that describes 
a steepest-descent trajectory towards the minimum of the canonical energy
functional (i.e. \eqref{F.eq} with $\mu=0$) 
\cite{Dalfovo1996}. This dynamics is not
unitary, and the correct
normalization to $N$  must be re-established at each time step. When the
stationary state is reached, $\varphi(x,\tau)=
e^{-\mu\tau}\varphi(x)$, the 
chemical potential is easily extracted from the required global renormalization
factor. Figure \ref{mu_q.fig} shows the
measured relative shift $\delta|\mu|/|\mu$ as function of $k_0$ for
various potential strengths ($N=100$). The analytical prediction
\eqref{deltamu_lattice.eq} is indeed found to be correct for $V_0\ll
|\mu|$. The inset shows the collapsed data obtained after division by
$V_0/|\mu|$, clearly featuring the interesting $k_0$-dependence, with the
maximal shift reached close to $k_0=1$.

%-------------------------------------------------------------------
\subsection{Expansion coefficients and small parameter}
\label{lattice_coeffs.sec} 

The expansion coefficients \eqref{alphaj.eq} read 
\begin{align} 
\alpha_0 & = - \frac{V_0}{2|\mu|}\sqrt{\frac{N}{6}}\frac{\pi}{2}
(1+k_0^2)\sech\frac{\pi k_0}{2},\\
\alpha_k &=  - \frac{V_0}{2|\mu|}\frac{\sqrt{\pi N}}{4} \frac{1+k^2-3
  k_0^2}{(1+k^2)^{3/2}(4+k^2)^{1/2}} \nonumber \\
 & \qquad \times \left[ \sech\tfrac{\pi}{2}(k+k_0)
  +\sech\tfrac{\pi}{2}(k-k_0)\right].
\end{align} 
All coefficients scale as $V_0/|\mu|$  by construction. 
As function of $k_0$, the bound-state coefficient
$\alpha_0 \sim k_0^2 e^{-\pi k_0/2}$ becomes exponentially small for a rapidly fluctuating potential
$k_0\gg1$ (cf.\ the qualitatively similar behaviour of the effective
potential \eqref{tildeVq.eq}) and thus does not contribute
substantially in this limit.  
Away from $|k|=k_0\gg 1$, also the coefficients $\alpha_k$ are 
exponentially small. However, the resonant
amplitudes $\alpha_{\pm k_0}\sim
k_0^{-2}$ are only algebraically small. We therefore conclude that the
modes $\pm k_0$ contribute dominantly, imprinting a pure sinusoidal
wave onto the soliton, with an overall weight scaling as 
\begin{equation} 
\frac{V_0}{|\mu|k_0^2} =  \frac{2V_0}{k^2}= \frac{V_0}{E_k}.
\end{equation} 
In this smoothing regime, completely analogous to the repulsive case 
 \cite{Sanchez-Palencia2006}, the healing length (or
chemical potential) drops out, and the small parameter of the
expansion rather is the ratio of potential amplitude to the kinetic 
energy $E_k$  
associated with the spatial lattice wave vector $k$. 

%-------------------------------------------------------------------
\subsection{Real-space response}

%---------------------------------------------
\begin{figure}
\begin{center}
\includegraphics[width=0.9\linewidth]{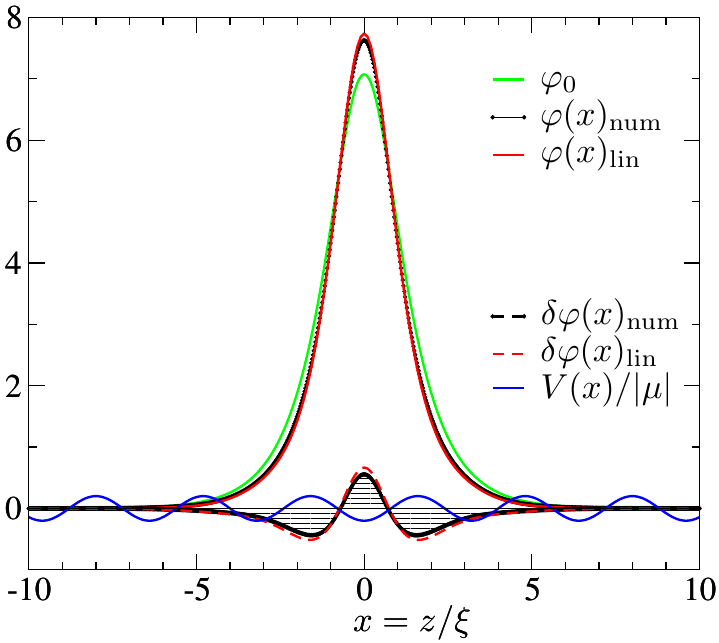}
\caption{(Color online) A soliton amplitude [green: $\varphi_0(x)$] containing  $N=100$
  particles and centered at $x=0$ is
  deformed by the lattice potential \eqref{Vlattice.eq}  with $V_0
  =0.2|\mu|$ and wave vector $k_0=5\pi/8$ [blue: $V(x)/|\mu|$]. The numerically
  calculated GP ground state [black: $\varphi(x)_\text{num}$]  
  is very well
  approximated by the linear response result [red: $\varphi(x)_\text{lin}$]
 given by eq.\
  \eqref{deltaphiN.eq}. The respective deviations $\delta\varphi(x)$ are
  also shown in dashed: the amplitude enhancement around the potential
  minimum at $x=0$ is clearly visible, just as the amplitude depression around
  the first potential maxima. 
}
\label{phi_lattice.fig}
\end{center}
\end{figure}
%-----------------------------------------------------

Unfortunately, even for the simple lattice potential
\eqref{Vlattice.eq} the integral \eqref{delphik.eq} over the kernel 
\eqref{kernel.eq} cannot be evaluated to simple closed form. However, it is
easily calculated numerically. 
Fig.\ \ref{phi_lattice.fig} shows how the unperturbed soliton ground state
$\varphi_0(x) = \sqrt{N/2}\sech x$ with $N=100$ particles is deformed by
a lattice potential with a rather large amplitude, $V_0=0.2|\mu|$ and
intermediate lattice wave vector, $k_0= 5\pi/8 \approx 1.9635$. 
The prediction from the linear-response theory matches the full result
quite well, considering that the perturbation is rather strong. 
To reach this agreement, it is essential  to take  into account the chemical potential
shift according to eq.\ \eqref{deltaphiN.eq}.   
The deviations $\delta\varphi(x)$ are also shown in dashed. 
One clearly observes a concentration of density around the potential
minimum, as well as density depressions around the first potential
maxima. 

%---------------------------------------------
\begin{figure}
\begin{center}
\includegraphics[width=\linewidth]{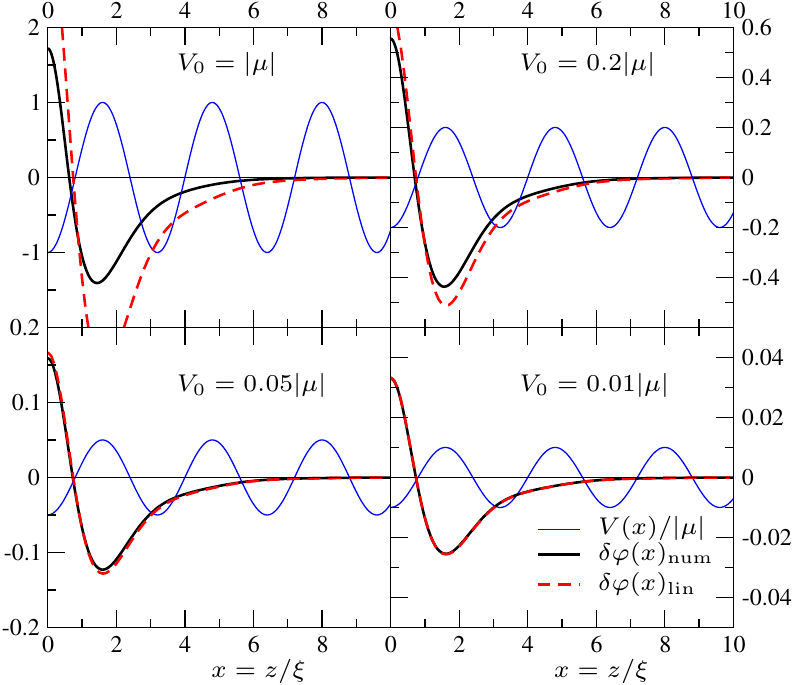}
\caption{(Color online) Deformation of the soliton shape by a lattice potential [blue: $V(x)/|\mu|$] of fixed wave
  vector $k_0=5\pi/8$ and various amplitudes $V_0$. The
  linear-response result,  eq.\
  \eqref{deltaphiN.eq} [dashed red: $\delta\varphi(x)_\text{lin}$] approaches the numerically
  obtained ground state deformation 
  [solid black: $\delta\varphi(x)_\text{num}$] as $V_0/|\mu|\to 0$. 
}
\label{delphi_v.fig}
\end{center}
\end{figure}
%-----------------------------------------------------

To check the validity and precision of the linear response, we plot
the deformation in direct comparison to the data from the numerical
solution,  together with the potential in Fig.\ \ref{delphi_v.fig} for
fixed intermediate wave vector $k_0=5\pi/8$ and various potential
strengths, $V_0/|\mu| \in\{ 1, 0.2, 0.05, 0.01\}$. By parity, we can
restrict the plots to $x\ge 0$. As expected, the linear
response approches the full solution  very
well as soon as $V_0\ll|\mu| $.

%---------------------------------------------
\begin{figure}
\begin{center}
\includegraphics[width=\linewidth]{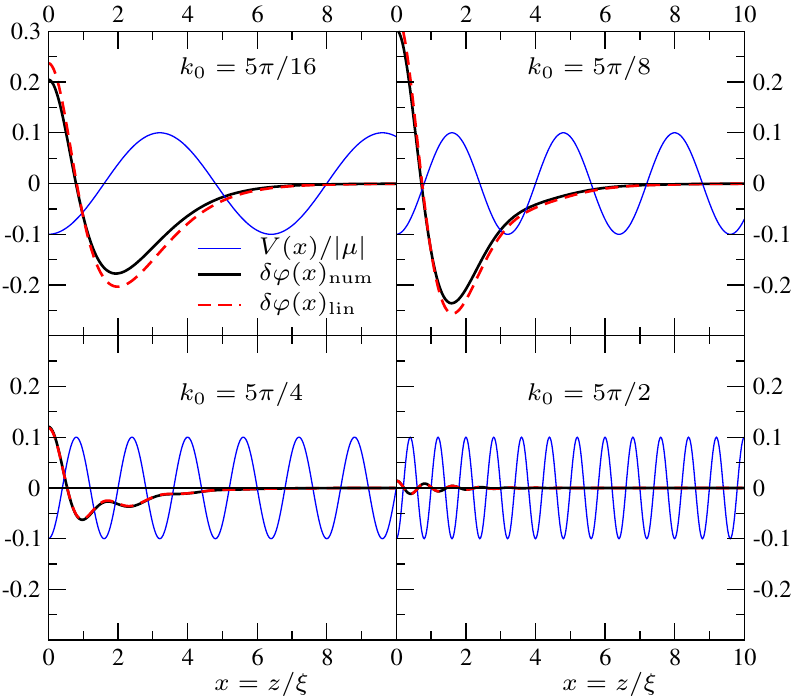}
\caption{(Color online) Deformation of the soliton shape by a lattice potential
  [blue: $V(x)/|\mu|$] of fixed  amplitude $V_0=0.1|\mu|$ and various
  wave  vectors $k_0$. The
  linear-response result eq.\
  \eqref{deltaphiN.eq} [dashed red:  $\delta\varphi(x)_\text{lin}$] approaches the numerically
  obtained ground state deformation 
  [black: $\delta\varphi(x)_\text{num}$] as $k_0\gg1$. 
}
\label{delphi_k.fig}
\end{center}
\end{figure}
%-----------------------------------------------------

Finally, Fig.\ \ref{delphi_k.fig} shows the same data for fixed 
 potential strength  $V_0 =0.1 |\mu|$ and various wave vectors 
$ k_0 \in 5\pi\{\frac{1}{16}, \frac{1}{8}, \frac{1}{4},\frac{1}{2}\}$.
Note that the vertical axis retains the same scaling in all four
plots. 
As expected from the discussion in Sec.\ \ref{lattice_coeffs.sec} above, the linear
response becomes increasingly accurate for larger $k$ vectors, as the
deformation itself becomes very small.  The
last set of data shows an almost pure sinusoidal deformation. Indeed, in the
limit $k_0\gg 1$, one can extract the leading contribution of \eqref{delphik.eq}  by partial
integration and finds 
\begin{equation} 
\delta \varphi(x)  = \frac{V_0}{E_k} \varphi_0(x) \cos k_0x   %\left[1+O\{ (V/E_k)^2\}\right] 
\end{equation} 
up to higher orders in $V_0/E_k = 2V_0/k^2$.

%-------------------------------------------------------------------
\section{Concluding remarks}
\label{conclusion.sec}

In this article, we elaborate on two important aspects of bright 
cold-atom solitons in inhomogeneous external potentials. Firstly, we
discuss the quantum dynamics of the soliton's center of mass in
a disordered, spatially correlated potential. The center-of-mass wave
function is predicted to show Anderson
localization with a localization length that can be estimated
using the Born approximation of perturbation theory and is 
shown to be in an experimentally relevant range. 

Secondly, we calculate the imprint of a weak external
potential on the soliton mean-field shape in the ground state. For
this, we determine the linear-response kernel, both for the wave function and the local
density, as well as the chemical potential shift. Amusingly, this can
be done rather easily using the supersymmetric quantum mechanics  of the
P\"oschl-Teller or $\sech^2$-potential well. 
Finally, a detailed
comparison to the numerically calculated ground-state solution is
presented for a pure
lattice potential with amplitude $V_0$ and wave vector $k$. The small
parameter of the linear response is $V_0/|\mu|$ in the
regime $k\xi \approx 1$ of a potential varying on the scale of the
soliton width. In the limit
$k\xi \to 0$, one recovers the case of harmonic confinement, studied
recently by Castin \cite{Castin2009}. Here, the small parameter for
the soliton deformation \eqref{delphik.eq}  is $V_0k^2\xi^2/2|\mu|=
(\omega_0/2|\mu|)^2$ in terms of the harmonic trapping
frequency $\omega_0$.   
In the smoothing limit $k\xi\gg1$ of a rapidly fluctuating potential,
the condensate amplitude shows a pure sinusoidal imprint, the small
parameter being $V_0/E_k$. 

Until now, we have discussed the center-of-mass quantum dynamics for fixed
shape on the one hand, and the static properties of the mean-field
ground state shape on the other. From the latter, we have learned that
the impact on the soliton shape is indeed small for small
$V_0/|\mu|$. In other words, the external potential cannot easily
excite the internal modes since these have a gapped
spectrum. Therefore, whenever the soliton moves slowly within a smooth
potential, it is slightly polarized, but only adiabatically and
reversibly (just as, say, an alkali atom is polarized in an optical dipole
potential).  Therefore, we can expect the center-of-mass quantum
dynamcis to be unharmed by shape excitations as long as the gap
$|\mu|$ remains large compare to all other energies.  

It must be kept in mind, however, that Anderson localization is an interference effect relying on
perfect phase coherence, at least over the time and distance needed
to observe it. Already very small decoherence can kill this
effect. For cold-atom solitons, the most dangerous sources
of decoherence arguably are scattering of background-gas atoms (cf.\ the analogous case of Fullerenes
\cite{Hornberger2003}) 
and three-body collisions, both of which can in principle be minimized
in the experiment. 
However, if the movement of a soliton inside a disordered
potential were to radiate permanent 
excitations, as discussed for the case of isolated point impurities in
\cite{Gredeskul1992}, then this would constitute an intrinsic
source of decoherence. The detailed study of such effects, well beyond the simple mean-field
estimate for the ground state presented here, is left for future research.

\begin{acknowledgements}
This work is supported by the National Research Foundation \& Ministry of
Education, Singapore. It was initiated in 
fiery discussions with K.\ Sacha, J.\ Zakrzewski and D.\ Delande 
at Laboratoire Kastler Brossel (Paris) 
while the author held a research fellowship from Mairie de Paris. 
Helpful input by B.\ Gr\'emaud  is 
gratefully acknowledged.  
\end{acknowledgements}

%\bibliography{soliton}   % name your BibTeX data base

\end{document}